\documentclass[twocolumn, eqsecnum, aps ]{revtex4}
\usepackage{graphicx}
\usepackage{dcolumn}

\begin{document}

\title{Electron Correlation Effects in Hyperfine Interactions in $^{45}$Sc and $^{89}$Y} 
\vspace{0.5cm}

\author{$^{1,2}$Bijaya K. Sahoo \protect \footnote[2] {E-mail: bijaya@iiap.res.in, B.K.Sahoo@gsi.de}, $^{1}$Thomas Beier, $^{2}$B. P. Das, $^{2}$R. K. Chaudhuri, $^{3}$Debashis Mukherjee \\
\vspace{0.3cm}
$^{1}${\it Atomphysik, \\ Gesellschaft f\"ur Schwerionenforschung mbH, Planckstra$\beta$e 1, 64291 Darmstadt, Germany}\\
$^{2}${\it Non-Accelerator Particle Physics Group,\\ Indian Institute of
Astrophysics, Bangalore-34, India}\\
$^{3}${\it Department of Physical Chemistry, \\Indian Association for Cultivation of Science, Calcutta-700 032, India}}
\date{Received date; Accepted date}
\vskip1.3cm
\begin{abstract}
\noindent
The relativistic coupled-cluster theory has been employed to calculate the magnetic dipole and electric quadrupole hyperfine structure constants for the stable isotopes $^{45}$Sc and $^{89}$Y. The role of electron correlation is found to be very important. The trend exhibited by these effects is rather different from previously studied single valence atomic systems. 
\end{abstract} 
\maketitle

\section{Introduction}
Electron correlation effects play a crucial role in the accurate determination of hyperfine struture constants \cite{lindgren}. We have calculated these constants for several atomic systems with single $s$ valence electrons \cite{sahoo01,sahoo02}  using the relativistic coupled-cluster (RCC) theory which takes into account the interplay of relativistic and correlation effects. Our recent work on Pb$^+$ which has a single $p$ valence electron \cite{sahoo03} reveals the peculiar behaviour of the electron correlation effects in the hyperfine constant of the $6p ^2P_{3/2}$ state. It would indeed be instructive to study how electron correlation effects influence systems with single $d$ valence electrons. 
In this paper we have carried out {\it ab initio} calculation of the magnetic dipole ($A$) and electric quadrupole ($B$) hyperfine structure constants for stable Scandium (Sc) and Yttrium (Y) which have $3d$ and $4d$ valence electrons respectively. Hyperfine structure constants for ground and first excited states for these two atoms have been measured using different methods \cite{fricke1,childs,siegfried,crawford,kuhn,brun}. 

\section{THEORY}
The hyperfine interaction Hamiltonian with magnetic dipole and electric quadrupole terms can be written as \cite{schwartz}
\begin{eqnarray}
{\it H_{hf}} = A \textbf{I.J} + B \left \{ \frac {3( \textbf{I.J})^2 + \frac {3}{2} \textbf{I.J} - I(I+1)J(J+1)}{2I(2I-1)J(2J-1)} \right\}, 
\end{eqnarray}
where $A$ and $B$ are the hyperfine structure constants and  {\bf I} and {\bf J} are the nuclear and electronic angular momenta respectively. In perturbation theory, the first order energy shift in the atomic state $|JM_J\rangle$ with $M_J$ as azimuthal quantum number corresponding to $J$ due to the above interaction Hamiltonian yields
\begin{eqnarray}
W_F = \frac {AK}{2} + \frac {B}{4} \left [\frac {\frac {3}{2}K(K+1)-2I(I+1)J(J+1)}{I(2I-1)J(2J-1)} \right],
\end{eqnarray}
where
\begin{eqnarray}
K= F(F+1)-J(J+1)-I(I+1)
\end{eqnarray}
with
\begin{eqnarray}
\textbf{F}=\textbf{I}+\textbf{J}
\end{eqnarray}
is the total angular momentum of the system.\\

The magnetic dipole hyperfine constant $A$ is given by \cite{cheng}
\begin{eqnarray}
A = \mu_N g_I \frac {\langle J || \textbf{T}^{(1)}||J\rangle}{\sqrt{J(J+1)(2J+1)}}, 
\end{eqnarray}
where $\textbf{T}^{(1)}=\sum t_q^{(1)} = \sum -ie \sqrt{8\pi/3} r^{-2} \alpha.\textbf{Y}_{1q}^{(0)}(\hat{r})$ with $\textbf{Y}_{kq}^{(\lambda)}$ represents the vector spherical harmonic, $g_I= \frac {\mu_I}{I}$ is the nuclear Land$\acute{e}$ g-factor with $\mu_I$ is nuclear dipole moment and $\mu_N=e\hbar/2m_pc$.\\

The electric quadrupole hyperfine constant ($B$) is given by \cite{cheng}
\begin{eqnarray}
B = 2eQ \left \{\frac {2J(2J-1)}{(2J+1)(2J+2)(2J+3)} \right \}^{1/2} \langle J|| \textbf{T}^{(2)}||J\rangle,
\end{eqnarray}
where $Q$ is the nuclear quadrupole moment and $\textbf{T}^{(2)}=\sum t_q^{(2)} = \sum -er^{-3} C_q^{(k)}(\hat{r})$.\\

The reduced matrix elements of the above operators in terms of single particle orbitals given by
\begin{eqnarray}
\langle \kappa_f || t_q^{(1)} || \kappa_i \rangle = - (\kappa_f+\kappa_i) \langle -\kappa_f || C^{(1)} || \kappa_i \rangle \nonumber \\ \int_0^{\infty} dr \frac {1}{r^2} (P_fQ_i+Q_fP_i)
\end{eqnarray}
and 
\begin{eqnarray}
\langle \kappa_f || t_q^{(2)} || \kappa_i \rangle = - \langle \kappa_f || C^{(2)} || \kappa_i \rangle \int_0^{\infty} dr \frac {1}{r^3} (P_fP_i+Q_fQ_i) \nonumber, \\
\end{eqnarray}
where $i$ and $f$ represent initial and final orbitals respectively and
\begin{eqnarray}
\langle \kappa_f || C^{(k)} || \kappa_i \rangle = (-1)^{j_f+1/2} \sqrt{(2j_f+1)(2j_i+1)} \nonumber \\ \left ( \matrix { j_f & k & j_i \cr 1/2 & 0 & -1/2 \cr } \right ) \pi(l_f,k,l_i) 
\end{eqnarray}
with the condition $\pi(l1,l2,l3)=1$ if $l1+l2+l3=even$, otherwise $zero$.

\section{METHOD OF CALCULATION : Relativistic Coupled Cluster Theory}
We have performed our calculations in the Dirac-Coulomb (DC) approximation which embodies the dominant relativistic and correlation effects. The Breit interaction which is about two orders of magnitude smaller than the Coulomb interaction has therefore been omitted in the present case as the systems are comparatively smaller in size.\\

The Dirac-Coulomb atomic Hamiltonian is given by
\begin{equation}
H = \sum_{j} c {\bf\alpha}.{\bf p}_{j} + (\beta -1) c^{2} + V_{nuc}(r_j) +
\sum_{j<l} \frac {1}{r_{jl}},
\end{equation}
where $\alpha$ and $\beta$ are the usual Dirac matrices and $V_{nuc}(r_j)$ is the potential at the site of the $j^{th}$ electron due to the atomic nucleus.
The rest mass energy of the electron is subtracted from the energy eigen values. We first solve the relativistic Hartree-Fock (Dirac-Fock (DF)) equations to obtain the single particle orbitals and their energies,
$$H_{DF} = \sum_{j} c {\bf\alpha}.{\bf p}_{j} + (\beta -1)c^{2} + V_{nuc}(r_j) + U_j$$.
The residual Coulomb interaction is given by 
\begin{equation}
V_{es} = \sum_{j<l} \frac {1}{r_{jl}} - \sum_{j} U_j .
\end{equation}
The single particle orbitals are obtained by solving the following equation self-consistently
\begin{equation}
(t_j + U_j ) |\phi_j\rangle = \epsilon_j|\phi_j\rangle, 
\end{equation} 
where
\begin{eqnarray}
t_j &=& c \alpha \cdot {\bf p}_j + (\beta - 1) c^2 +V_{nuc}(r_j)\nonumber
\end{eqnarray}
and
\begin{eqnarray}
U_j |\phi_j(\vec r_1)\rangle &=& \sum_{a=1}^{occ} \langle \phi_a(\vec r_2) |\frac{1}{r_{12}}|\phi_a(\vec r_2) \rangle |\phi_j(\vec r_1) \rangle \nonumber \\
& - & \langle \phi_a(\vec r_2) |\frac{1}{r_{12}}|\phi_j(\vec r_2) \rangle |\phi_a(\vec r_1) \rangle\nonumber
\end{eqnarray}
\noindent
{\it occ} represents total number of occupied orbitals.\\

The single particle relativistic orbitals can be expressed as
$$
|\phi_j(r)\rangle = \frac {1}{r} \left (
                        \matrix {
                         P_j(r) |\chi_{\kappa_j m_j} \rangle \cr
                         Q_j(r) |\chi_{-\kappa_j m_j} \rangle \cr 
                         }
                        \right ),
$$
where $P_j(r)$ and $Q_j(r)$ are the radial part of the large and small components respectively and $|\chi_{\kappa_j m_j}\rangle$ and $|\chi_{-\kappa_j m_j}\rangle$ are their respective spin angular momentum components. $\epsilon_j$'s are the single particle energies.\\

We have employed the RCC to incorporate correlation effects among electrons due to the residual Coulomb interaction. In this approach the exact atomic wavefunction for the closed-shell system can be expressed as \cite{lindgren} 
\begin{equation}
|\Psi_{CC}\rangle = e^T |\Phi\rangle,
\end{equation}
where T is the core electron excitation operator and we call it as closed-shell RCC operator. $|\Phi\rangle$ is the above closed-shell determinantal state built out of the Dirac-Fock single particle orbitals.\\

In the closed-shell coupled-cluster theory one starts with the equation
\begin{equation}
H e^T |\Phi\rangle = E e^T |\Phi\rangle.
\end{equation}
The energy and amplitude determining equations are
\begin{equation}
\langle \Phi^K |\overline{H}|\Phi \rangle = E \delta_{0,K}, 
\end{equation}
where $\overline{H} = e^{-T} H e^{T}$, $|\Phi^K\rangle$ is a determinantal state with K = 0,1,2.... representing the reference state and excited determinantal states.
We have considered all possible non-linear terms in T- operator for its amplitude determining equations. \\

Goldstone \cite{lindgren,szabo01,bartlett} and angular momentum diagrammatic \cite{lindgren,edmonds01} techniques are used for evaluating different radial integrals and angular factors. The normal ordered Hamiltonian is defined as
\begin{equation}
H_N \equiv H - \langle \Phi |H| \Phi \rangle =  H - E_{DF},
\end{equation}
where $E_{DF}=\langle \Phi|H|\Phi\rangle$.\\

We have truncated our wavefunction expansion at the level of singles and doubles (CCSD) and all possible non-linear terms have been included in the above equation.
First we evaluate wavefunction for the closed shell system using the above RCC approach and then append the corresponding valence electron 
 using the open shell RCC (OSCC) method \cite{lindgren,debasish}.\\
  
The new reference state of the open-shell system with one valence electron {\it v} can be expressed as \cite{debasish} 
\begin{equation}
|\Phi_v\rangle\ \equiv\ a_v^{\dag}|\Phi\rangle,
\end{equation}
where $a_v^{\dag}$ is the particle creation operator. The exact atomic states are defined now, using the Fock-space OSCC method, as
\begin{equation}
|\Psi_v\rangle = e^T\{e^{S_v}\}|\Phi_v\rangle,
\end{equation}
where $S_v$ is the valence excitation operator and we call it as open-shell RCC operator. Since systems under consideration can be treated as with one valence electron, the $S_v$- operator exponential series naturally truncates at the linear term, i.e. the open-shell wavefunction has the form
\begin{equation}
|\Psi_v\rangle = e^T\{1+S_v\}|\Phi_v\rangle,
\end{equation}
where
\begin{eqnarray}
S_v\ =\ S_{1v} + S_{2v} = \sum_{p \ne v}a_p^+a_v s_v^p + \frac {1}{2}\sum_{bpq}a_p^+a_q^+a_ba_v s^{pq}_{vb} \nonumber
\end{eqnarray}
and
\begin{eqnarray}
 S_{1v} = \sum_{p \ne v} a_p^+a_v s_v^p \nonumber \\
 S_{2v} = \frac {1}{2}\sum_{bpq}a_p^+a_q^+a_ba_v s^{pq}_{vb}
\end{eqnarray}
with $s_v^p$ and $s^{pq}_{vb}$ are the cluster amplitudes corresponding to single and double excitations involving the valence electron.\\

In the next step, we include approximate triple excitations by contracting the two-body operator ($V_{es}$) and the double excitation operators ($T_2,S_{2v}$) in the following way \cite{kaldor01} 
\begin{equation}
S_{vbc}^{pqr}\ =\ \frac{\widehat{V_{es}T_2}+\widehat{V_{es}S_{2v}}}{\epsilon_v+\epsilon_b+\epsilon_c-\epsilon_p
-\epsilon_q-\epsilon_r},
\end{equation}
where $\epsilon_i$ is the orbital energy of the {\it i'th} orbital. Note that we use notations a,b,c..., p,q,... and i,j,... for the core (occupied), particle (unoccupied) and general orbitals respectively.\\

The equations for the open-shell cluster amplitudes are determined from
\begin{eqnarray}
\langle \Phi_v|\overline{H_N}\{1+S_v\}|\Phi_v\rangle = \Delta E(v)  
\end{eqnarray}
and
\begin{eqnarray}
\langle \Phi_v^{*}|\overline{H_N}\{1+S_v\}|\Phi_v\rangle = -\Delta E(v) \langle \Phi_v^{*}|
\{S_v\}|\Phi_v\rangle,
\end{eqnarray}
where $\Delta E(v)$ is the electron attachment energy which is equal to the negative of the ionisation potential for the valence electron {\it v}.\\

The expectation value for a general one particle operator in a given valence electron ($v$) state can be expressed in coupled-cluster theory as
\begin{eqnarray}
 \langle O \rangle_v  &=& \frac {\langle\Psi_v | O | \Psi_v \rangle} {\langle\Psi_v|\Psi_v\rangle} \nonumber \\
 &=& \frac {\langle \Psi_v | O | \Psi_v \rangle} {1+N_v} \nonumber \\
 &=& \frac {\langle \Phi_v | \{1+S_v^{\dagger}\} e^{T^{\dagger}} O e^T \{1 + S_v\} | \Phi_v\rangle } {1+N_v}\\
 &=& \frac {\langle \Phi_v | \{1+S_v^{\dagger}\} \overline{O} \{1 + S_v\} | \Phi_v\rangle } {1+N_v},
\end{eqnarray}
\noindent
where the normalization term for the {\it v$^{th}$} orbital is obtained from
\noindent
\begin{eqnarray}
N_v &=& \langle \Phi_v | S_v^{\dagger} [e^{T^{\dagger}} e^T] + S_v^{\dagger} [e^{T^{\dagger}} e^T] S_v + [e^{T^{\dagger}} e^T] S_v | \Phi_v\rangle\nonumber \\
 &=& \langle \Phi_v | S_v^{\dagger} \overline{n_v} + S_v^{\dagger} \overline{n_v} S_v + \overline{n_v} S_v^{\dagger} | \Phi_v\rangle
\end{eqnarray}
with
\noindent
\begin{equation}
\overline{O}=(e^{T^{\dagger}} O e^T)_{f.c.} + (e^{T^{\dagger}} O e^T)_{o.b.} + (e^{T^{\dagger}} O e^T)_{t.b.} + ....
\end{equation}
and
\noindent
\begin{equation}
\overline{n_v} = (e^{T^{\dagger}} e^T)_{f.c.} + (e^{T^{\dagger}} e^T)_{o.b.} + (e^{T^{\dagger}} e^T)_{t.b.} + ...
\end{equation}

The f.c., o.b., t.b.,..etc abbreviations are used for the fully contracted, effective one-body, effective two-body ...etc terms respectively \cite{sahoo01}. Terms containing only up to effective three-body diagrams will contribute to both the numerator and the denominator. The fully contracted terms are excluded on the basis of the linked-diagram theorem \cite{lindgren} in the evaluation of the $\overline{O}$ and $\overline{N}$. All the one-body terms have been taken into account as their contribution to the correlation effects is the largest. The dominant parts of the two-body terms have also been computed \cite{sahoo01}. Finally, these terms are contracted with $S_v^{\dagger}$ and $S_v$ operators.\\

Contributions from the normalization factor have been determined in the following way 
\begin{eqnarray}
Norm = \langle \Psi_v | O | \Psi_v \rangle \{ \frac {1}{1+N_v} - 1 \}.
\end{eqnarray}

\section{Results and Discussion}
\begin{table*}[t]
\begin{ruledtabular}
\begin{center}
\caption{Magnetic dipole ($A$) and electric quadrupole ($B$) hyperfine structure constants in $MHz$ of ground state and first excited state for Sc and Y }
\begin{tabular}{lcccc|lcccc}
 & Scandium & & & & & Yttrium & & & \\
States & Theory (Ours) & & Experiment \cite{fricke,childs} & & States & Theory (Ours) & & Experiment \cite{siegfried,fricke1,crawford,kuhn,brun} & \\
 & $A$  & $B$ & $A$  & $B$ &  & $A$ & $B$  & $A$  & $B$ \\
\hline
 $3d ^2D_{3/2}$ & 268.67  & 27.45  & 269.556(1) & 26.346(4) & $4d ^2D_{3/2}$ & 57.34  & 23.37 & 57.217(15) & -  \\
 $3d ^2D_{5/2}$ & 110.75 & 38.56 & 109.032(1) & 37.31(10) & $4d ^2D_{5/2}$ & 29.22 & 31.75 & 28.749(30) & - \\
\end{tabular}
\end{center}
\end{ruledtabular}
\end{table*}

\begin{table}[t]
\caption{Contributions from different coupled-cluster terms to the magnetic dipole
hyperfine structure constant ($A$) in $MHz$, where $cc$ stands for the complex conjugate part of the corresponding terms}
\begin{ruledtabular}
\begin{tabular}{l|cc|cc}
 & Scandium & & Yttrium & \\ 
\hline
CCSD(T) & & &  &  \\
terms &$3d ^2D_{3/2}$ & $3d ^2D_{5/2}$ & $4d ^2D_{3/2}$ & $4d ^2D_{5/2}$ \\
\hline
 & & & & \\
 $O$ (DF) & 235.93 & 100.37 & 56.06 & 23.24 \\
$\overline{O} - O $ & 4.82 & 2.41 & 2.37 & 1.04 \\
$\overline{O} S_{1v} + cc $ & 40.02 & 15.64 & 7.62 & 3.11 \\
$\overline{O} S_{2v} + cc $ & -20.09 & -22.93 & -8.27 & -5.08 \\
$S_{1v}^{\dagger} \overline{O} S_{1v}$ & 1.85  & 0.67 & 0.26 & 0.11 \\
$S_{1v}^{\dagger} \overline{O} S_{2v} + cc $ & 1.21 & -2.30  & 0.32 & -0.54 \\
$S_{2v}^{\dagger} \overline{O} S_{2v} + cc $ & 25.79 & 25.59 & 3.62 & 9.45 \\
\hline\\
\multicolumn{5}{c}{Important effective two-body terms of $\overline{O}$}\\
\hline
 & & & & \\
$S_{2v}^{\dagger} O T_1 + cc $ & -9.07 & -3.77 & -2.05 & -0.85\\
$S_{2v}^{\dagger} O T_2 + cc $ & 0.00 & 0.00 & -0.32 & -0.02 \\
 $Norm.$ & -10.94 & -4.61 & -2.38 & -1.19\\
\hline\\
 Total & 268.67 & 110.75 & 57.34 & 29.22 \\
\end{tabular}
\end{ruledtabular}
\label{tab:front3}
\end{table}
\begin{table}[t]
\caption{Contributions from different coupled-cluster terms to the magnetic dipole
hyperfine structure constant ($B$) in $MHz$, where $cc$ stands for the complex conjugate part of the corresponding terms}
\begin{ruledtabular}
\begin{tabular}{l|cc|cc}
 & Scandium & & Yttrium & \\ 
\hline
CCSD(T) & & &  &  \\
terms &$3d ^2D_{3/2}$ & $3d ^2D_{5/2}$ & $4d ^2D_{3/2}$ & $4d ^2D_{5/2}$ \\
\hline
 & & & & \\
 $O$ (DF) & 23.34 & 32.76 & 16.01 & 21.78 \\
$\overline{O} - O $ & 0.78 & 0.84 & 0.69 & 0.76 \\
$\overline{O} S_{1v} + cc $ & 4.05 & 5.53 & 2.18 & 2.94 \\
$\overline{O} S_{2v} + cc $ & -0.97 & -1.18 & 3.90 & 5.78 \\
$S_{1v}^{\dagger} \overline{O} S_{1v}$ & 0.18  & 0.24 & 0.07 & 0.11 \\
$S_{1v}^{\dagger} \overline{O} S_{2v} + cc $ & -0.24 & -0.33  & 0.06 & 0.99 \\
$S_{2v}^{\dagger} \overline{O} S_{2v} + cc $ & 2.74 & 3.99 & 2.03 & 2.44 \\
\hline\\
\multicolumn{5}{c}{Important effective two-body terms of $\overline{O}$}\\
\hline
 & & & & \\
$S_{2v}^{\dagger} O T_1 + cc $ & -0.91 & -1.16 & -0.59 & -0.79\\
$S_{2v}^{\dagger} O T_2 + cc $ & 0.00 & 0.00 & -0.01 & -0.10\\
 $Norm.$ & -1.14 & -6.7 & -0.97 & -1.30 \\
\hline\\
 Total & 27.45 & 38.56 & 23.37 & 31.75 \\
\end{tabular}
\end{ruledtabular}
\label{tab:front3}
\end{table}

We have used Gaussian type orbitals (GTOs) for the construction of single particle orbitals of the Dirac-Fock wavefunction($|\Phi \rangle$), whose expression is given by \cite{rajat1} 
\begin{equation}
F_{i,k}^{(L/S)}(r) = \sum_i c_i^{(L/S)} r^k e^{-\alpha_i r^2}
\end{equation}
with k=0,1,2,... for $s$, $p$, $d$,$\cdots$ respectively. The function $F_{i,k}^{(L/S)}(r)$ stands for the large (L) and small (S) components of the Dirac wavefunction. $ c_i^{(L/S)}$ is the expansion coefficient of the corresponding large and small components respectively. The kinetic balance condition \cite{stanton01} has been imposed between the large and small components of the GTOs. For the exponents, the even tempering condition \cite{raffenetti}
\begin{equation}
\alpha_i = \alpha_{i-1} \beta , \hspace*{2cm} i=1,\cdots,N
\end{equation}
has been applied. Here, $N$ stands for the total number of basis functions for a specific symmetry.
All orbitals
are generated on a grid using a two-parameter Fermi nuclear distribution
approximation given  by
\begin{equation}
\rho = \frac {\rho_0} {1 + e^{(r-c)/a}},
\end{equation}
where $\rho_0$ is the average nuclear density, the parameter 'c' is the {\it half-charge radius,} and 'a' is related to
the {\it skin thickness} which is defined as the interval of the nuclear
thickness which the nuclear charge density falls from near one to near zero.\\

\begin{figure}
\label{fig:goldstone}
\includegraphics[width=9.0cm]{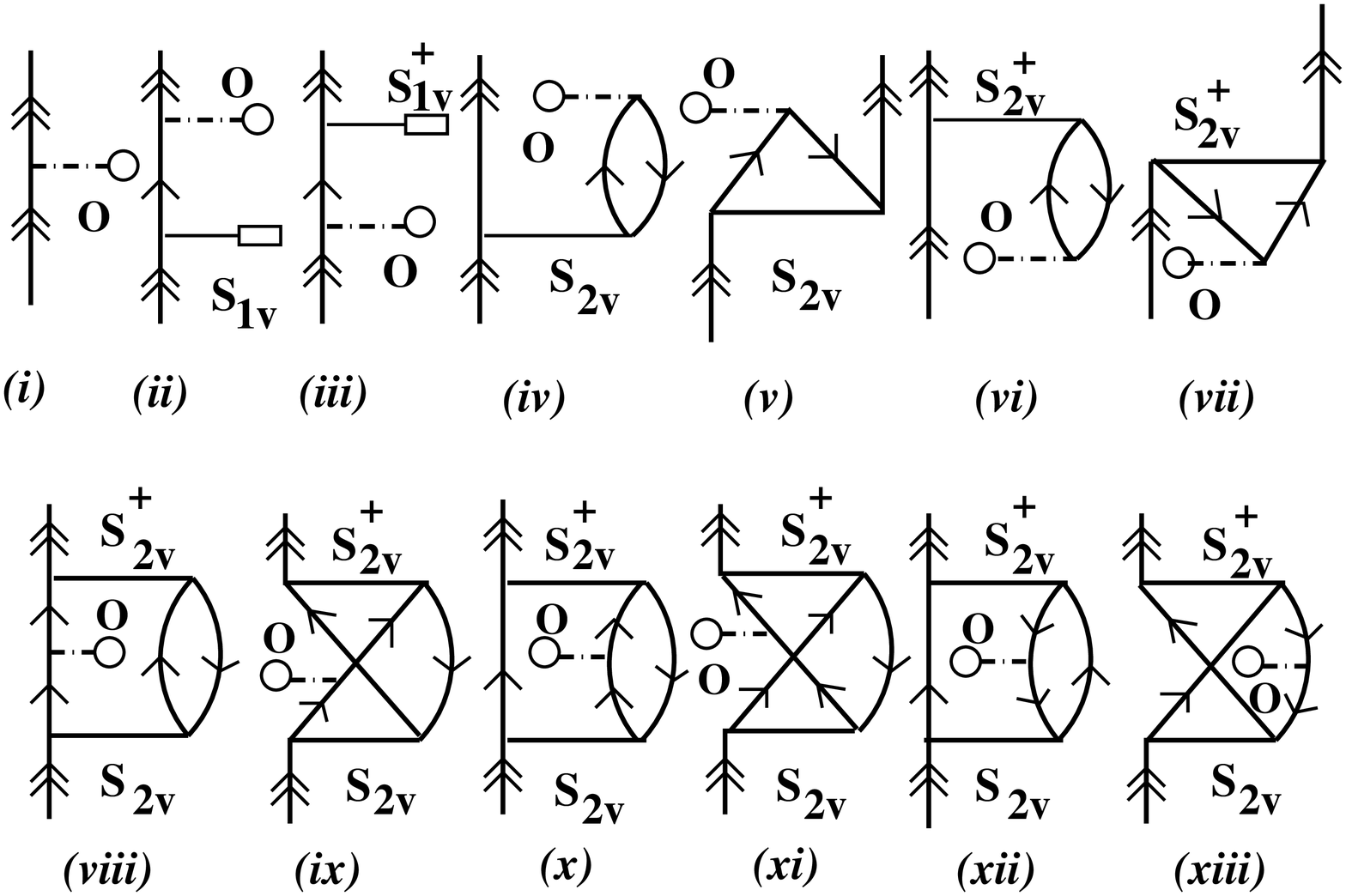}
\caption{Goldstone diagrams representing hyperfine structure calculations}
\end{figure}

In the present calculation, we have used $\alpha_{\circ}$ = 0.00525 and $\beta$ = 2.73 for all symmetries in both the systems. We have considered 35$s_{1/2}$, 30$p_{1/2}$, 30$p_{3/2}$, 30$d_{3/2}$, 30$d_{5/2}$, 20$f_{5/2}$ and 20$f_{7/2}$  and 38$s_{1/2}$, 35$p_{1/2}$, 35$p_{3/2}$, 35$d_{3/2}$, 35$d_{5/2}$, 20$f_{5/2}$ and 20$f_{7/2}$ GTOs to obtain Dirac-Fock wavefunction in Sc and Y respectively. All core electrons have been excited in the present calculations.\\

We have calculated both $A$ and $B$ constants for $3d ^2D_{3/2}$ and $3d ^2D_{5/2}$ states for Sc and  $4d ^2D_{3/2}$ and $4d ^2D_{5/2}$ states for Y. We have used $g_I=1.359$ and $Q=0.22$ for Sc and $g_I=0.3268$ and $Q=0.125$ for Y respectively. All calculated results with corresponding experimental results are presented in table I.\\

In table II and table III, we present contributions from different RCC terms for $A$ and $B$ values respectively. The first term $O$ represents the DF value and is given in FIG. 1 (i). The second term $\overline{O} - O $ is the contribution from the correlation between the core electros. $OS_{1v}$ (FIG. 1 (ii)+(iii)) and $OS_{2v}$ (FIG. 1 ((iv)+(v)+(vi)+(vii))) which represent pair-correlation and core-polarization effects respectively make important contributions for both the atoms. It is interesting to note that the contributions of $S_{2v}^{\dagger} O S_{2v}$ (FIG. 1 ((viii)+(ix)+(x)+(xi)+(xii)+(xiii))) is far more significant for Sc and Y than atomic systems with single $s$ and $p$ valence electrons. One of the strengths of coupled-cluster theory is to account for such intricate correlation effects to all orders in the residual Coulomb interaction. The two-body $\overline{O}$ and  normalization terms make non-negligible contributions.

\section{Conclusion}
We have performed calculations of the hyperfine constants $A$ and $B$ for Sc and Y using RCC theory and high lighted the importance of electron correlation. Our results are in good agreement with experiments. This indeed demonstrates the power of RCC theory to capture the interplay of relativistic and correlation effects for atomic systems with single $d$ valence electrons.

\section{Acknowledgment}
\noindent
This work was carried out under the DAAD project of the first three authors. A part of the computations were carried out on the Teraflop Supercomputer, C-DAC, Bangalore, India.

\end{document}